\documentclass[lettersize,journal]{IEEEtran}

\usepackage{amsmath,amsfonts}
\usepackage{mathtools}
\usepackage{algorithmic}
\usepackage{graphicx}
\usepackage{textcomp}
\usepackage{hyperref}
\usepackage{xcolor}
\usepackage{braket}
\usepackage{tabularx}
\usepackage{pgfplots}
\usepackage{pgf}
\usepackage{booktabs}
\usepackage{glossaries}
\usepackage{listingsutf8}
\usepackage{array}
\usepackage{multirow}
\usepackage{orcidlink}
\usepackage{balance}

\usepackage{subcaption}
\usepackage{todonotes}
\usepackage[free-standing-units]{siunitx}

\usepackage[utf8]{inputenc}

\newacronym{gnn}{GNN}{Graph Neural Network}
\newacronym{ttl}{TTL}{Time to Live}
\newacronym{lstm}{LSTM}{Long Short-Term Memory}

\AtBeginDocument{%
  }

\newboolean{arxiv}
\setboolean{arxiv}{true} 
\ifthenelse{\boolean{arxiv}}{
    \usepackage{eso-pic}
    \AddToShipoutPictureBG*{
                    \AtPageLowerLeft{%
                        \raisebox{2\baselineskip}{
\makebox[\paperwidth]{\fbox{\begin{minipage}{0.85\paperwidth}\scriptsize
                                        {This work has been submitted to the IEEE for possible publication. Copyright may be transferred without notice, after which this version may no longer be accessible.}
                                    \end{minipage}}}}%
                    }
    }
}{} 

\begin{document}
\bstctlcite{IEEEexample:BSTcontrol}

\markboth{Preprint submitted to the IEEE}{SatQNet: Satellite-assisted Quantum Network Entanglement Routing Using Directed Line Graph Neural Networks}

\title{SatQNet: Satellite-assisted Quantum Network Entanglement Routing Using Directed Line Graph Neural Networks}

\newcommand{\hrefnoborder}[2]{%
  \begingroup
  \hypersetup{pdfborder={0 0 0}}%
  \href{#1}{#2}%
  \endgroup
}

\author{
\IEEEauthorblockN{Tobias Meuser\IEEEauthorrefmark{1}\orcidlink{0000-0002-2008-5932}, Jannis Weil\IEEEauthorrefmark{1}\IEEEauthorrefmark{2}\orcidlink{0000-0001-5439-9131}, Aninda Lahiri\IEEEauthorrefmark{3}\orcidlink{0000-0003-4833-8046}, Marius Paraschiv\IEEEauthorrefmark{3}\orcidlink{0000-0002-4761-9842}}\\
\IEEEauthorblockA{
\IEEEauthorrefmark{1}Communication Networks Lab, Technical University of Darmstadt, \hrefnoborder{mailto:tobias.meuser@kom.tu-darmstadt.de}{tobias.meuser@kom.tu-darmstadt.de}\\
\IEEEauthorrefmark{2}Institute of Communication Technology, Leibniz University Hannover, \hrefnoborder{mailto:jannis.weil@ikt.uni-hannover.de}{jannis.weil@ikt.uni-hannover.de}\\
\IEEEauthorrefmark{3}Quantum Information Group, IMDEA Networks, \{\hrefnoborder{mailto:aninda.lahiri@imdea.org}{aninda.lahiri}, \hrefnoborder{mailto:marius.paraschiv@imdea.org}{marius.paraschiv}\}@imdea.org
}
\vspace{-0.45cm}
}

\maketitle

\begin{abstract} 
Quantum networks are expected to become a key enabler for interconnecting quantum devices.
In contrast to classical communication networks, however, information transfer in quantum networks is usually restricted to short distances due to physical constraints of entanglement distribution.
Satellites can extend entanglement distribution over long distances, but routing in such networks is challenging because satellite motion and stochastic link generation create a highly dynamic quantum topology.
Existing routing methods often rely on global topology information that quickly becomes outdated due to delays in the classical control plane, while decentralized methods typically act on incomplete local information.

We propose SatQNet, a reinforcement learning approach for entanglement routing in satellite-assisted quantum networks that can be decentralized at runtime.
Its key innovation is an edge-centric directed line graph neural network that performs local message passing on directed edge embeddings, enabling it to better capture link properties in high-degree and time-varying topologies.
By exchanging messages with neighboring repeaters, SatQNet learns a local graph representation at runtime that supports agents in establishing high-fidelity end-to-end entanglements.
Trained on random graphs, SatQNet outperforms heuristic and learning-based approaches across diverse settings, including a real-world European backbone topology, and generalizes to unseen topologies without retraining.

\end{abstract}


\section{Introduction}

To unlock the full potential of quantum computing and quantum sensing, it is essential to develop robust quantum communication networks that enable quantum systems to interact and share information over long distances.
Quantum communication harnesses phenomena such as entanglement to transmit quantum information securely, for example, via quantum teleportation.
Quantum networks~\cite{li2023, singh2021, li2021} are essential not only for scaling up quantum computing capabilities but also for enabling new technologies in quantum sensing and metrology.
Entanglement distribution across a network allows for enhanced precision in measurements and synchronization tasks, which can revolutionize fields such as navigation and time measurement~\cite{8930960, K_m_r_2014}.
A global quantum network, or quantum internet, would facilitate new forms of secure communication and information processing~\cite{azuma2023}, fundamentally changing the landscape of information technology~\cite{sandilya2021, illiano2022, chehimi2022}.

Quantum networks can be envisioned as an overlay atop classical communication infrastructures like fiber optics, where entanglement is distributed from initial elementary links (entangled pairs established between neighboring nodes) through entanglement swapping~\cite{zukowski1994}.
In this architecture, entanglement swapping performed by quantum repeaters~\cite{biswas2023} plays a crucial role in extending the range of entanglement distribution, enabling entangled states to be shared between nodes that are not directly connected.

While the literature on entanglement distribution in terrestrial quantum networks with stationary repeaters is extensive~\cite{shi2020, zeng2022, avis2023, sen2023}, few studies address the complexities of satellite-based quantum network routing~\cite{biswas2023, pirandola2021, chang2024a}.
Terrestrial fiber-optic networks exhibit attenuation, causing an exponential decay of entangled photon transmission rates over long distances~\cite{krutyanskiy2019, yu2020}.
This limits the entanglement distribution to a few hundred kilometers without repeaters, leading to elementary links that only cover comparatively small distances.
Even with repeaters, the maximum achievable distance depends strongly on the length of each elementary link, as each swap operation introduces additional noise to the entangled state.
Satellite-based networks offer a promising avenue for global quantum communication by bridging large distances that are impractical for ground-based links.
These networks comprise both ground-based repeaters and a fleet of satellites capable of performing entanglement swapping and equipped with local quantum memories for storing qubits.

Implementing satellite-based quantum networks introduces unique challenges~\cite{wang2014, vallone2014, agnesi2019, wang2022} to both the network itself and entanglement routing.
Aspelmeyer et al.~\cite{aspelmeyer2003} provide a detailed list of hardware requirements for satellite-assisted long-distance quantum communication.
The effects of atmospheric loss and noise significantly impact the absorption and transmission of photons between satellites and ground stations.
Satellite orbital dynamics cause the lengths of links to ground stations to vary continuously, leading to a time-dependent network topology.
This poses additional challenges to entanglement routing, as it further increases the already high dynamics of quantum networks.
In particular, approaches relying on global information about the quantum network, as presented in \cite{li2022,xiong2023}, scale poorly with the size of satellite-assisted quantum networks, as the necessary information can only be gathered with significant latency.
In addition, even local approaches often rely on knowledge about the underlying physical topology~\cite{chakraborty2019}, which is often assumed to be static over time, limiting their applicability to satellite-assisted quantum networks.
Although learning-based approaches offer a possible solution to entanglement routing, many approaches are trained and evaluated on a single topology or require a global view of the quantum network, making them unsuitable for satellite-assisted networks~\cite{le2022,abreu2024}.
Even approaches that generalize across topologies tend to make assumptions about the structure of the underlying network, e.g., by limiting the maximum degree of quantum repeaters~\cite{meuser2025} or only generalizing to relatively small networks~\cite{geyer2018,weil2024}, which also limit their applicability to satellite-assisted quantum networks.
In this work, we present SatQNet, a learning-based approach that relies on learned graph representations to make routing decisions in highly dynamic satellite-assisted quantum networks.
The key technical novelty of SatQNet is the use of a directed line GNN.
In contrast to related work~\cite{geyer2018,weil2024,meuser2025}, which commonly relies on node embeddings that encode information about adjacent links only indirectly, SatQNet maintains separate embeddings for each link at each node and therefore better captures link-specific dynamics in satellite-assisted quantum networks.

Our contributions are as follows:
\begin{itemize}
    \item We model time-varying ground-to-satellite and inter-satellite link success probabilities and formulate a satellite-aware entanglement-routing problem.
    \item We propose SatQNet, a learning-based decentralized routing policy that leverages a directed line \gls*{gnn} and is trained with reinforcement learning.
    \item We perform an extensive evaluation across synthetic and real backbone topologies with statistical testing.
\end{itemize}

The structure of the paper is as follows: Section~\ref{related_works} reviews the relevant literature in the field, while Section~\ref{sec:model} covers our quantum network model and key concepts in quantum information. In Section~\ref{sec:link-success-probabilities}, we model the link success probabilities for ground-to-satellite and inter-satellite links, followed by a description of our reinforcement learning approach in Section~\ref{sec:rl}. Section~\ref{evaluation} presents the training process and discusses the results, along with a performance comparison of the proposed method against selected heuristics and other learning-based approaches. Finally, the paper concludes with a summary in Section~\ref{conclusion}.

\section{Related Work}
\label{related_works}

Several comprehensive review articles have surveyed the recent progress of quantum networks from various perspectives~\cite{li2023, singh2021, li2021, sandilya2021, illiano2022, chehimi2022}.
Significant advancements have been made in the area of entanglement routing across networks with diverse topologies. Research in entanglement routing has focused on both bipartite entangled states~\cite{shi2020, zeng2022, zhang2022, hahn2019, inesta2023} and multipartite entangled states~\cite{mannalath2023, sutcliffe2023, avis2023, sen2023}.

The literature presents a variety of routing methodologies, including multi-path routing~\cite{pant2019, nguyen2022, zhang2022, li2021a}, which seeks to establish multiple parallel entanglement links between source and destination nodes concurrently. These links can subsequently undergo entanglement purification procedures~\cite{victora2023a}. The integration of multi-path routing with time-multiplexed quantum repeaters is explored in~\cite{milligen2025}, while issues related to resource under-utilization and the consequent loss of entangled pairs are addressed in~\cite{pouryousef2023}. Alternatively, some approaches utilize virtual graphs constructed from entangled links that are generated on demand~\cite{schoute2016, chakraborty2019}.

Graph-based techniques also play a role in entanglement distribution, utilizing operations like subgraph complementation~\cite{sen2023}. These methods are particularly advantageous in the analysis of graph states~\cite{adcock2020}, where local complementations facilitate the identification of classes of locally equivalent graphs. Moreover, in simple network topologies such as linear chains of repeaters, optimal entanglement routing schemes can be determined~\cite{dai2020, gu2024}.
While most research in quantum networking has concentrated on terrestrial quantum networks, the incorporation of satellites has only recently gained attention~\cite{biswas2023, deforgesdeparny2023, agnesi2018, gu2025, wei2025}. Experimentally, recent studies have focused on tracking methods~\cite{wang2014}, timing precision, and various implementation challenges~\cite{vallone2014, agnesi2019, wang2022}. Theoretically, there have been developments in secure quantum communications within satellite-based quantum networks~\cite{mishra2024}, as well as the derivation of various theoretical bounds~\cite{pirandola2021}.

Reinforcement learning has long been explored for pathfinding in classical networks~\cite{boyan1993}.
Many of these methods rely on centralized control~\cite{almasan2022}, which limit their scalability.
Decentralized approaches offer better scalability but suffer from limited network observability, hindering their performance~\cite{schneider2021}.
This challenge is particularly pronounced in dynamic environments like quantum networks, where frequent topology changes necessitate current network information for optimal decision-making.
To achieve generalizability in pathfinding tasks, \glspl*{gnn} have been employed in several studies~\cite{rusek2020}, although these typically depend on a centralized network view.
Recent advancements have introduced techniques that utilize  message passing GNNs to enable generalizability in pathfinding~\cite{geyer2018, weil2024,meuser2025}.
However, these methods operate based on node information, which is suboptimal for an edge-centric routing task, and their generalizability is constrained to networks with certain properties.

\section{Preliminaries and System Model}
\label{sec:model}

We model the quantum communication network based on an underlying optical network, referred to as the physical network, which connects quantum repeaters.
Over this physical network, neighboring quantum repeaters establish elementary links, which form the basic resources for entanglement routing.

To describe these resources formally, we work with qubits and, more generally, density operators $\rho \succeq 0$, $\mathrm{Tr}(\rho)=1$, which capture both pure states and statistical mixtures.
For a bipartite system $\rho_{AB} \in \mathcal{H}_A \otimes \mathcal{H}_B$, the reduced state is $\rho_A = \mathrm{Tr}_B(\rho_{AB})$~\cite{nielsen2010}.
The four Bell states
\begin{align}
    \ket{\Phi^\pm} = \tfrac{1}{\sqrt{2}}\!\left(\ket{00}\pm\ket{11}\right), \qquad
    \ket{\Psi^\pm} = \tfrac{1}{\sqrt{2}}\!\left(\ket{01}\pm\ket{10}\right)
\end{align}
are the canonical maximally entangled resources that serve as elementary links in our setting.
Local Paulis permute them, so many noise processes reduce to random Pauli errors on an ideal Bell pair.
Noise and loss are modeled as completely positive and trace preserving (CPTP) maps $\mathcal{E}$.
The primary instance used here is the depolarizing channel, whose action on a Bell pair yields a \emph{Werner state}
\begin{equation}
\begin{aligned}
    W_F
    &=  F \left| \Phi^+ \right\rangle \left\langle \Phi^+ \right|
     + \frac{1 - F}{3} \Big(
        \left| \Phi^- \right\rangle \left\langle \Phi^- \right| \\
    &\qquad\qquad
      + \left| \Psi^+ \right\rangle \left\langle \Psi^+ \right|
      + \left| \Psi^- \right\rangle \left\langle \Psi^- \right|
     \Big).
\end{aligned}
\label{wernerstate}
\end{equation}
In \eqref{wernerstate}, $W_F$ is a uniform mixture of all four Bell states weighted by a fidelity parameter $F$, defined as
\begin{equation}
    F(\rho) \;=\; \bra{\Phi^{+}}\,\rho\,\ket{\Phi^{+}} \in [0,1].
\end{equation}
The fidelity represents the overlap of the considered state $\rho$ with a target state, in this case the pure Bell state $\ket{\Phi^{+}}$. Throughout the paper, the fidelity value associated with an \emph{elementary} entangled pair between two repeaters serves as the primary scalar measure of link quality.

\ifthenelse{\boolean{arxiv}}{
\begingroup
\definecolor{tabblue}{RGB}{31, 119, 180}
\definecolor{tabred}{RGB}{214, 39, 40}
\definecolor{tabgreen}{RGB}{44, 160, 44}
\definecolor{taborange}{RGB}{255, 127, 14}

\definecolor{probhigh}{RGB}{33, 102, 172}     
\definecolor{probmedhigh}{RGB}{146, 197, 222} 
\definecolor{probmed}{RGB}{244, 165, 130}     
\definecolor{problow}{RGB}{178, 24, 43}       

\tikzset{
    node distance=1.5cm,
    node/.style={circle, draw, minimum size=.75cm},
    connection/.style={thick, gray},
    connection-wireless/.style={thick, gray, dotted},
    prob-high/.style={thick, probhigh},
    prob-medhigh/.style={thick, probmedhigh},
    prob-med/.style={thick, probmed},
    prob-low/.style={thick, problow},
    prob-wireless-high/.style={thick, dashed, probhigh},
    prob-wireless-medhigh/.style={thick, dashed, probmedhigh},
    prob-wireless-med/.style={thick, dashed, probmed},
    prob-wireless-low/.style={thick, dashed, problow}
}

\def\posSOnex{1.0}
\def\posSOney{2.5}
\def\posSTwox{3.0}
\def\posSTwoy{2.5}

\def\posSOnexMoved{2.0}
\def\posSOneyMoved{2.5}
\def\posSTwoxMoved{4.4}
\def\posSTwoyMoved{2.5}

\def\antennaoffset{0.52}

\begin{figure}[tb]
    \centering
    \subfloat[Initial topology\label{subfig:initial}]{
        \includegraphics{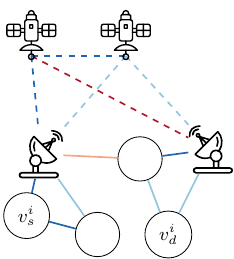}
    }
    \hspace{0.25cm}
    \subfloat[Topology after movement\label{subfig:moved}]{
        \includegraphics{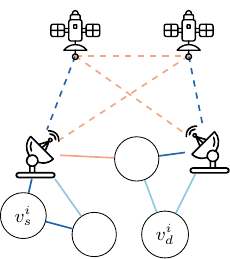}
    }
    \caption{Simplified example of a satellite-assisted quantum network. The quantum repeaters are depicted as nodes, connected via fiber links. Dedicated ground stations (satellite dishes) have both fiber and satellite connections. Link colors represent entanglement generation probability, with blue indicating higher probability (shorter links) transitioning to red for lower probability (longer links).}\label{fig:topology}
\end{figure}
\endgroup
}{

}
\subsection{Quantum repeaters}
To establish end-to-end entanglement over longer distances, elementary links must be combined through entanglement swapping at intermediate quantum repeaters.
In our model, we consider two types of repeaters, as shown in Fig.~\ref{fig:topology}: stationary repeaters and mobile repeaters.
\textit{Stationary repeaters} are fixed nodes located at specific sites, such as small data centers, and are interconnected via optical fibers.
These fiber connections are stable and remain constant during network operation.
A subset of these stationary repeaters is equipped with ground-to-air communication interfaces, enabling them to establish connections with mobile repeaters.
\textit{Mobile repeaters}, exemplified by satellites, can communicate both with stationary repeaters and with other mobile repeaters through air-based optical links, whose properties may differ depending on the involved nodes.
Because mobile repeaters change their positions over time, the availability and quality of these links vary dynamically.

\subsection{Generation and Decoherence of Elementary Links}\label{subsec:model_links}
Elementary links are created with a given probability and form the basis of the quantum topology.
This probability depends on the distance and connection type between nodes $v_1$ and $v_2$.
For fiber-based links, this probability decays exponentially with distance, while the behavior of air-based links is more complex and is modeled in Sec.~\ref{sec:link-success-probabilities}.
The probability of establishing elementary links via fiber is constant over time, whereas the probability for air-based links changes dynamically due to node mobility.
This behavior can be observed by comparing Fig.~\ref{subfig:initial} with Fig.~\ref{subfig:moved}.

After their creation, elementary links are stored in quantum memories and degrade over time due to decoherence~\cite{nielsen2010, breuer2002}.
We rely on the decay model developed in \cite{meuser2025} based on the measurements of \cite{stas2022}, which is shown in \eqref{eq:fidelity_decay}.
\begin{equation}\label{eq:fidelity_decay}
    F(t) = \left(F_0 - F_{B}\right) \cdot \exp\left[-\left(\frac{t}{T_{2,n_{\text{dec}}}}\right)^k\right] + F_B
\end{equation}

Together, the currently available elementary links define the quantum topology and provide the resources consumed by entanglement swapping.

\subsection{Entanglement Swapping}
\label{sec:entanglement-swap}
\emph{Entanglement swapping} enables two nodes without a direct entangled link to become entangled via an intermediate node~\cite{zukowski1994, pan1998}.
Operationally, swapping consumes two input pairs and produces one output pair (possibly only probabilistically, depending on the hardware)~\cite{sun2017}.
Imperfect local operations can be incorporated as additional local noise that reduces the resulting fidelity.
In this work, we simulate the swapping procedure in Qiskit~\cite{javadiabhari2024}.

\subsection{Routing in Quantum Networks}\label{subsec:routing}
In our quantum network model, nodes may request end-to-end entanglement with other nodes in the network.
We denote such a request by a source-destination pair $(v^i_s, v^i_d)$.
As in classical communication networks, a route for a source-destination pair is a sequence of neighboring nodes that starts at the source node $v^i_s$ and ends at the destination node $v^i_d$.

Using the available elementary links, the goal is to establish end-to-end entanglement for each active source-destination pair.
This is achieved by performing a sequence of entanglement swaps along the selected path.
The objective is to maximize the quality of the resulting end-to-end entanglement, measured primarily through its fidelity, while also maximizing the number of successful end-to-end entanglements.

Once a path is planned, the repeaters along that path perform swap operations, thereby successively extending the entanglement.
The quality of each repeater along the path introduces additional noise during the swaps, reducing the fidelity of the resulting entanglement.
Consequently, the fidelity of an end-to-end entanglement depends not only on the availability of elementary links, but also on their fidelities and on the quality of the repeaters performing the swap operations.

Thus, finding high-fidelity routes in satellite-assisted quantum networks requires knowledge of the current quantum topology, i.e., the quality and availability of elementary links.
During the routing process, the quantum topology may change frequently, and routing approaches must adapt accordingly.

\subsection{Quantum network operation}
During the operation of our quantum network model, all neighboring repeaters attempt to establish as many elementary links as possible between them.
At the same time, paths can be planned to establish end-to-end entanglements between all active source-destination pairs.
We adopt a continuous entanglement generation approach, in which elementary links are generated continuously while the routing is performed.
This continuous generation of entangled links ensures that the network maintains available quantum resources throughout the routing operation, enhancing the efficiency and reliability of entanglement distribution.
While this increases the number of available elementary links, it also makes the quantum topology more dynamic.
Once a new elementary link is created, the maximum fidelity of any elementary link between two repeaters will increase.
In addition, links without any available elementary link, which would otherwise be considered unusable, might generate a new elementary link, effectively reactivating the link from the perspective of the quantum topology.

\section{Optical Link Model for Ground-to-satellite and Inter-satellite Links}
\label{sec:link-success-probabilities}

\begin{table*}[t]
  \centering
  \footnotesize
  \setlength{\tabcolsep}{4pt}
  \begin{tabularx}{\textwidth}{l X >{\raggedright\arraybackslash}p{0.3\textwidth} >{\raggedright\arraybackslash}p{0.3\textwidth}}
    \toprule
    Parameter & Meaning & Ground-to-satellite & Inter-satellite \\
    \midrule
    $R_E$ & Earth radius & $6371\,\mathrm{km}$ & $6371\,\mathrm{km}$ \\
    $h_s$ & Satellite altitude & $500$--$1200\,\mathrm{km}$ (LEO) & $500$--$600\,\mathrm{km}$ \\
    $\lambda$ & Optical wavelength & $850\,\mathrm{nm}$ (e.g.\ Micius downlink) & $850\,\mathrm{nm}$ or $1550\,\mathrm{nm}$ \\
    $D_T$ & Tx aperture diameter & $0.30\,\mathrm{m}$ (satellite) \cite{liao2017a, lu2022a} & $0.10\,\mathrm{m}$ (typical ISL terminal) \cite{kaushal2017a} \\
    $D_R$ & Rx aperture diameter & $1.0\,\mathrm{m}$ (ground) \cite{liao2017a, lu2022a} & $0.10\,\mathrm{m}$ \\
    $\theta_{\text{div}}$ & Full beam divergence & $\sim 10\,\mu\mathrm{rad}$ (GS/SG) \cite{liao2017a, lu2022a} & $20$--$40\,\mu\mathrm{rad}$ (SS) \cite{kaushal2017a} \\
    $T_{\text{zenith}}$ & Zenith transmittance & $\approx 0.8$ (clear night at $850\,\mathrm{nm}$) \cite{liao2017a, lu2022a} & $1$ \\
    $\tau_{\text{zenith}}$ & Zenith optical depth & $\approx 0.22$ ($T_{\text{zenith}}\approx 0.8$) & $0$ \\
    $\sigma_p$ & RMS pointing jitter per terminal & $0.2$--$0.5\,\mu\mathrm{rad}$ (fine tracking) \cite{lu2022a} & $5$--$10\,\mu\mathrm{rad}$ (SS) \cite{kaushal2017a, farid2007a} \\
    $\eta_{\text{opt,tx}}$ & Tx internal optical efficiency & $0.5$--$0.8$ & $0.5$--$0.8$ \\
    $\eta_{\text{opt,rx}}$ & Rx internal optical efficiency & $\approx 0.16$ (ground telescope + coupling) \cite{liao2017a} & $0.5$--$0.8$ \\
    $\eta_{\text{det}}$ & Detector efficiency & $\approx 0.5$ (Si SPADs at $850\,\mathrm{nm}$) \cite{liao2017a} & $0.5$--$0.9$ (e.g.\ SNSPDs at $1550\,\mathrm{nm}$) \\
    $\eta_{\text{source}}$ & Source efficiency (into the mode) & $0.3$--$0.7$ (protocol dependent) & $0.3$--$0.7$ (protocol dependent) \\
    $h_{\min}$ & Grazing altitude threshold in \eqref{eq:visibility} & $20\,\mathrm{km}$ & $20\,\mathrm{km}$ \\
    \bottomrule
  \end{tabularx}
  \caption{Representative parameter values for ground-to-satellite and inter-satellite links. Values are consistent with reported satellite quantum key distribution (QKD) demonstrations and free-space optical (FSO) link budgets~\cite{liao2017a, lu2022a, kaushal2017a, farid2007a, khatri2021a}.}
  \label{tab:link-params}  
\end{table*}

In the time-evolving graph representation of the network, each edge corresponds to a potential \emph{elementary quantum link}.
We associate with each edge a time-dependent \emph{single-photon success probability} $P_{\text{link}}(u,v,t)$, defined as the probability that a photon emitted by the local source into the spatial mode of the edge $(u,v)$ at time $t$ is successfully detected at the remote node after propagating through the channel.

We distinguish two types of links:
(i) ground--satellite or satellite--ground links (GS/SG), where one endpoint is a terrestrial
station and the other is a satellite;
(ii) inter-satellite links (SS), where both endpoints are satellites connected through a
vacuum free-space optical (FSO) channel.
We model both SS and GS/SG communications based on established physical approaches and environmental constraints. For additional details the reader is also directed to Refs.~\cite{liao2017a, lu2022a, kaushal2017a}.

These link probabilities are used as edge weights by the higher-level learning and routing modules:
edges with large $P_{\text{link}}(u,v,t)$ are usable for entanglement distribution,
whereas edges with small $P_{\text{link}}(u,v,t)$ are effectively unavailable.

\subsubsection*{Geometry and airmass for ground--satellite links}

Let the Earth be modeled as a sphere of radius $R_E$, and let the ground station be located at
geodetic latitude $\phi_g$, longitude $\lambda_g$. A satellite at altitude $h_s$ has orbital
radius $r_s = R_E + h_s$ and geodetic coordinates $(\phi_s(t),\lambda_s(t))$ at time $t$.
The central angle $\gamma(t)$ between the ground station and the satellite sub-point is
\begin{equation}
  \cos\gamma(t)
  =
  \sin\phi_g\sin\phi_s(t)
  + \cos\phi_g\cos\phi_s(t)\cos\!\big(\lambda_s(t)-\lambda_g\big).
  \label{eq:central-angle}
\end{equation}
The corresponding ground--satellite slant range is
\begin{equation}
  L_{\text{GS}}(t)
  =
  \sqrt{r_s^2 + R_E^2 - 2R_E r_s \cos\gamma(t)}.
  \label{eq:slant-range-gs}
\end{equation}
The elevation angle $\mathrm{El}(t)$, i.e.\ the angle between the line-of-sight and the local
horizontal at the ground station, is
\begin{equation}
  \tan\!\big(\mathrm{El}(t)\big)
  = \frac{\cos\gamma(t) - (R_E/r_s)}{\sin\gamma(t)},
  \label{eq:elevation}
\end{equation}
and the zenith angle is $\zeta(t) = \tfrac{\pi}{2}-\mathrm{El}(t)$. In simulations we typically
impose a minimum elevation $\mathrm{El}(t) \ge \mathrm{El}_{\min}$ (e.g.\ $20^\circ$); below this
threshold we set the link probability to zero.

The atmospheric transmittance for a photon traversing the atmosphere at zenith angle $\zeta$ is
modeled by a Beer--Lambert law applied to a slant path:
\begin{equation}
  \eta_{\text{atm}}(\zeta)
  =
  \exp\!\big(-\tau_{\text{zenith}}\sec\zeta\big)
  =
  T_{\text{zenith}}^{\sec\zeta},
  \label{eq:eta-atm}
\end{equation}
where $\tau_{\text{zenith}}$ is the zenith optical depth at the signal wavelength and
$T_{\text{zenith}} = e^{-\tau_{\text{zenith}}}$ is the corresponding vertical transmissivity.
For clear mid-latitude nights around $\lambda\approx 850\,\mathrm{nm}$ one finds
$T_{\text{zenith}}\approx 0.8$ (i.e.\ $\tau_{\text{zenith}}\approx 0.22$)
\cite{liao2017a, lu2022a, kaushal2017a}.

\subsubsection*{Diffraction-limited geometric collection}

We model the optical beam in both GS/SG and SS links as a fundamental Gaussian mode. Let
$D_T$ and $D_R$ be the transmitter and receiver aperture diameters, and let $a_R = D_R/2$ be
the receiver radius. Denote by $\theta_{\text{div}}$ the (far-field) $1/e^2$ \emph{half-angle}
beam divergence and by $L$ the distance between transmitter and receiver.

In the far-field regime, the beam radius at distance $L$ is
\begin{equation}
  w(L) \simeq \theta_{\text{div}} L,
\end{equation}
and the fraction of power intercepted by the receiver aperture (the diffraction/geometric
transmissivity) is
\begin{equation}
  \eta_{\text{diff}}(L)
  =
  1 - \exp\!\left(-\frac{2a_R^2}{w(L)^2}\right).
  \label{eq:eta-diff-general}
\end{equation}
If we additionally assume diffraction-limited transmit optics with Gaussian waist
$w_0 \approx D_T/2$, then $\theta_{\text{div}} = \lambda/(\pi w_0) \approx 2\lambda/(\pi D_T)$ and
\eqref{eq:eta-diff-general} can be written explicitly as
\begin{equation}
  \eta_{\text{diff}}(L)
  =
  1 - \exp\!\left(
        -\frac{\pi^2 D_T^2 D_R^2}{8\lambda^2 L^2}
      \right).
  \label{eq:eta-diff-closed}
\end{equation}
\eqref{eq:eta-diff-closed} makes explicit the $1/L^2$ scaling inside the exponential
governing the geometric collection loss. In practice, the effective $\theta_{\text{div}}$ may be
larger than the diffraction limit because of intentional beam expansion and residual optical 
aberrations; in that case it is more robust to treat $\theta_{\text{div}}$ as an independent
hardware parameter and use \eqref{eq:eta-diff-general} directly
\cite{liao2017a, lu2022a, kaushal2017a, khatri2021a}.

\subsubsection*{Pointing statistics}

Residual pointing errors due to mechanical jitter and finite-bandwidth tracking cause additional
attenuation. For a single terminal we model the instantaneous coupling loss as
\begin{equation}
  \Lambda_p(\theta) \approx \exp\!\big(-G\theta^2\big),
  \quad
  G \approx \frac{8}{\theta_{\text{div}}^2},
\end{equation}
where $\theta$ is the instantaneous radial pointing error angle, and $G$ is a ``pointing stiffness'' parameter that depends only on the beam divergence \cite{kaushal2017a,farid2007a}.
If the elevation and azimuth pointing errors are independent zero-mean Gaussians with variance $\sigma_p^2$, the radial error $\theta$ is Rayleigh-distributed.
Averaging $\Lambda_p(\theta)$ over this distribution yields the mean pointing efficiency of a single terminal
\begin{equation}
  \bar\eta_{\text{point}}
  =
  \frac{1}{1 + 2G\sigma_p^2}
  =
  \frac{1}{1 + 16\sigma_p^2/\theta_{\text{div}}^2}.
  \label{eq:eta-point-single}
\end{equation}
For an inter-satellite link, both transmitter and receiver terminals are subject to independent
jitter, so the total mean pointing efficiency is approximately the product of two such factors,
\begin{equation}
  \bar\eta_{\text{point,SS}}
  \approx \bigl(\bar\eta_{\text{point}}\bigr)^2.
  \label{eq:eta-point-ss}
\end{equation}

\subsubsection*{Satellite--satellite visibility}

For a pair of satellites $i$ and $j$ with position vectors $\mathbf{r}_i(t)$ and $\mathbf{r}_j(t)$
in an Earth-centered frame, the inter-satellite distance is
\begin{equation}
  L_{ij}(t) = \bigl\|\mathbf{r}_i(t) - \mathbf{r}_j(t)\bigr\|.
\end{equation}
We introduce a binary visibility function $I_{\text{vis}}(i,j,t)$ that accounts for Earth
occultation. Let $\mathbf{d}(t) = \mathbf{r}_j(t) - \mathbf{r}_i(t)$ and define
\begin{align}
  s^\ast(t)
  &= \mathrm{clip}_{[0,1]}\!\left(
      -\frac{\mathbf{r}_i(t)\cdot \mathbf{d}(t)}{\lVert \mathbf{d}(t)\rVert^2}
    \right),
    \label{eq:s-star} \\[0.3em]
  r_{\min}(t)
  &= \bigl\lVert \mathbf{r}_i(t) + s^\ast(t)\,\mathbf{d}(t)\bigr\rVert.
    \label{eq:r-min}
\end{align}
as the minimum distance of the line segment between the two satellites to the Earth's center. We use the clip function here to force the $s(t)$ parameter into the $[0,1]$ range (so $s=0$ at satellite i, $s=1$ at satellite j, and $0 < s < 1$ for intermediary points). The clip function $\mathrm{clip}_{[0,1]}(x)$ is $0$ below the $[0,1]$ interval, $x$ inside the interval and $1$ above. 

We then set
\begin{equation}
  I_{\text{vis}}(i,j,t)
  =
  \begin{cases}
    1, & r_{\min}(t) > R_E + h_{\min},\\[0.3em]
    0, & \text{otherwise},
  \end{cases}
  \label{eq:visibility}
\end{equation}
where $h_{\min}$ is a grazing altitude threshold (we take $h_{\min}\approx 20\,\mathrm{km}$)
to ensure that the optical path does not cross the dense lower atmosphere
\cite{khatri2021a, kaushal2017a}.

\subsubsection*{Ground--satellite (GS/SG) link probability}

For a GS/SG channel we group all hardware efficiencies into a single factor
\begin{equation}
  \eta_{\text{HW,GS}}
  =
  \eta_{\text{source}}\,
  \eta_{\text{opt,tx}}\,
  \eta_{\text{opt,rx}}\,
  \eta_{\text{det}}\,
  \bar\eta_{\text{point,GS}},
\end{equation}
where
$\eta_{\text{source}}$ is the source efficiency (probability that an ``attempt'' injects a photon
into the correct spatial mode),
$\eta_{\text{opt,tx}}$ and $\eta_{\text{opt,rx}}$ are transmit and receive internal optical
efficiencies (including telescope and fiber coupling),
$\eta_{\text{det}}$ is the single-photon detector efficiency, and
$\bar\eta_{\text{point,GS}}$ is the mean pointing efficiency of the GS/SG link
(often close to unity for well-engineered tracking systems).

The per-photon success probability of a GS/SG link at time $t$ is then
\begin{equation}
  P_{\text{GS}}(t)
  =
  \eta_{\text{HW,GS}}\,
  \eta_{\text{atm}}\bigl(\zeta(t)\bigr)\,
  \eta_{\text{diff}}\bigl(L_{\text{GS}}(t)\bigr)
  \label{eq:P-gs}
\end{equation}
with $\eta_{\text{atm}}$ defined in \eqref{eq:eta-atm} and $\eta_{\text{diff}}$ in either
\eqref{eq:eta-diff-general} or \eqref{eq:eta-diff-closed}. This expression is symmetric between
downlink (satellite~$\rightarrow$~ground) and uplink (ground~$\rightarrow$~satellite); in practice
one may use different hardware parameters in $\eta_{\text{HW,GS}}$ for the two directions
(e.g.\ to reflect stronger turbulence and higher background noise in uplinks
\cite{liao2017a, lu2022a, kaushal2017a}).

For an entanglement-based protocol where, say, one photon of each pair is transmitted along the
GS/SG link and the other photon is detected locally at the source station, the elementary
entanglement success probability per attempt is obtained by multiplying $P_{\text{GS}}(t)$ by the
local detection probability at the source node.

\subsubsection*{Satellite--satellite (SS) link probability}

For an SS link between satellites $i$ and $j$, there is no atmospheric extinction
($\eta_{\text{atm}}\approx 1$) and both terminals jitter. We again group all non-geometric
efficiencies into
\begin{equation}
  \eta_{\text{HW,SS}}
  =
  \eta_{\text{source}}\,
  \eta_{\text{opt,tx}}\,
  \eta_{\text{opt,rx}}\,
  \eta_{\text{det}}\,
  \bar\eta_{\text{point,SS}},
\end{equation}
where $\bar\eta_{\text{point,SS}}$ is given by \eqref{eq:eta-point-ss}.
The per-photon SS success probability is then
\begin{equation}
    P_{\text{SS}}(i,j,t)
    =
    I_{\text{vis}}(i,j,t)\,
    \eta_{\text{HW,SS}}\,
    \eta_{\text{diff}}\bigl(L_{ij}(t)\bigr)
  \label{eq:P-ss}
\end{equation}
with $I_{\text{vis}}(i,j,t)$ from \eqref{eq:visibility} and $\eta_{\text{diff}}$ from
\eqref{eq:eta-diff-general} or \eqref{eq:eta-diff-closed}. As before, in entanglement-based
protocols the elementary entanglement success probability per attempt is obtained by multiplying
$P_{\text{SS}}(i,j,t)$ with the local detection probabilities at the relevant nodes.

Table~\ref{tab:link-params} summarizes recommended default values for the main parameters of \eqref{eq:P-gs} and \eqref{eq:P-ss}. The GS/SG values are chosen to be representative of the
Micius satellite downlink and similar experimental demonstrations
\cite{liao2017a, lu2022a, kaushal2017a}, while the SS values are motivated by
existing and proposed free-space optical inter-satellite terminals
\cite{kaushal2017a, farid2007a, khatri2021a}.

In the simulations one can either use the detailed decomposition in
\eqref{eq:P-gs}--\eqref{eq:P-ss}, or pre-collapse the hardware factors into single scalars
$\eta_{\text{HW,GS}}$ and $\eta_{\text{HW,SS}}$ calibrated from a more detailed link budget or
experimental measurements. The resulting $P_{\text{GS}}(t)$ and $P_{\text{SS}}(i,j,t)$ then form
the time-dependent edge weights for the effective network graph used by the routing and
reinforcement-learning components.

\section{SatQNet: Reinforcement Learning for Satellite-assisted Quantum Routing}
\label{sec:rl}

We propose to address the challenges of satellite-based quantum routing like dynamic link properties and high node degrees, as presented in Section~\ref{sec:model}, with deep reinforcement learning.
In particular, \glspl*{gnn} allow reinforcement learning agents to generalize over previously unseen topologies and network conditions~\cite{weil2024}, making them a suitable choice for the highly dynamic topologies of quantum networks.
In the following, we provide a short introduction to combining reinforcement learning with \glspl*{gnn}, followed by the modeling of the agents and nodes in the network.

\subsection{Generalizable Reinforcement Learning through \glspl*{gnn}\label{subsec:gnn_architecture}}

Reinforcement learning considers the interaction of an agent with an environment.
The agent observes the environment and selects an action based on the received observation. 
For each action, the agent receives a numerical reward that quantifies the quality of the selected action.
The goal is to maximize the long-term rewards and thereby optimally solve a given task.

In deep reinforcement learning, the agent's policy function that defines the mapping from observations to actions is represented by a neural network.
Deep reinforcement learning is agnostic to the neural network architecture, but the choice can inherently limit generalizability.
In particular, neural network architectures with fixed input and output dimensions cannot process graphs of arbitrary size and order.

Our main requirement for satellite-based quantum networks is that the approach should, at least conceptually, be able to generalize over any network topology.
As monitoring information in the network can be modeled in graph form, \glspl*{gnn} are a natural choice.
However, while technically compatible with any graph, feedforward message-passing \glspl*{gnn} have a fixed number of layers, which limits the extent to which information is propagated through the graph.
This limits the maximum supported path length between source and destination nodes.
Weil et al.~\cite{weil2024} leverage a recurrent \gls*{gnn} to continuously propagate information through the graph and thereby support arbitrary path lengths, but their approach is limited to graphs with a fixed number of nodes and a uniform node degree.
The extension of Meuser et al.~\cite{meuser2025} supports an arbitrary number of nodes, as the observations are independent of the graph order, but is still limited by a maximum node degree.
Geyer et al.~\cite{geyer2018} also use a recurrent \gls*{gnn} and propose modeling network interfaces as nodes in addition to routers.
This allows for unified processing of node and interface features and renders the approach applicable to graphs with arbitrary maximum node degree.
However, the use of one-hot encoded node identifiers as input features limit their approach to graphs of fixed order.

Our proposed approach supports graphs of arbitrary order, size, and maximum node degree.
We train a single model on a set of graphs and then apply it to arbitrary unseen graphs.

\subsubsection{Recurrent GNN fundamentals}
Let $G=(V,E)$ be a graph with nodes~$V$, undirected edges~$E$, and time-dependent input features $x^v_t,\, x^e_t$ for each node $v\in V$ and edge $e\in E$ at discrete time $t \in \{0,\, 1,\,2,\, \dots\}$.
This graph represents the satellite-based quantum network.
Analogous to related work~\cite{meuser2025}, we leverage a recurrent \gls*{gnn} where each node $v\in V$ manages a \emph{node embedding} $h^v_t$ that is updated as
\newcommand{\gnnagg}{\text{aggregate}}
\newcommand{\gnnupdate}{\text{update}}
\newcommand{\gnnencode}{\text{encode}}
\begin{equation}
\begin{aligned}
    \hat{h}^v_t &= \gnnencode \left(h^v_t, x^v_t\right)\\
    M^v_t &= \gnnagg \left(\left\{ (\hat{h}^v_t, x^{\{v,w\}}_t, \hat{h}^w_t) \mid \{v, w\} \in E\right\}\right)\\
    h^v_{t+1} &= \gnnupdate \left(\hat{h}^v_t,\, M^v_t\right),
\end{aligned}
\label{eq:message-passing}
\end{equation}

where \gnnencode, \gnnagg{}, and \gnnupdate{} are arbitrary differentiable functions.
The function \gnnencode{} encodes information from the node features into the node embedding, \gnnagg{} aggregates features from adjacent edges and the node embeddings of neighboring nodes, and \gnnupdate{} updates the node embedding with the aggregated information.
By repeated execution of \eqref{eq:message-passing}, node embeddings eventually propagate through the entire graph, which enables distant nodes to communicate.
We assume that the edge features $x^{\{u, w\}}$ are known to both nodes $u,\, w$ connected by the edge.
Note that \eqref{eq:message-passing} only requires nodes to exchange node embeddings with their direct neighbors, which implies that its execution can be distributed.

\subsubsection{Support of dynamic node degrees}\label{subsubsec:dynamic_node_degrees}
Conceptually, the above \gls*{gnn} is applicable to any graph as long as the individual input features $x^v_t,\, x^e_t$ do not depend on the order of the graph, the size of the graph, or the maximum node degree.
An agent~$i$ located at a quantum router $v\in V$ at time $t$ leverages local information available at this router to make a routing decision.
To select one of the neighbor nodes $a^i_t \in \mathcal{N}(v) \coloneq \left\{w \mid \{v, w\} \in E\right\}$ at time $t$ independent of the node degree, the agent's action space must resize dynamically.
Inspired by early works on reinforcement learning for routing~\cite{boyan1993}, we achieve this by processing each neighbor separately as
\begin{equation}
\begin{aligned}
    a^i_t = \text{argmax}_{w\in \mathcal{N}(v)} Q(o^i_t,\, h^v_{t+1},\, x^{\{v,w\}}_t,\, h^w_t),
\end{aligned}
\label{eq:action-selection}
\end{equation}

where $Q$ can be any differentiable function that returns a scalar for each neighbor $w$ based on the information available to agent $i$ and node $v$ at time $t$.
Here, $o^i_t$ refers to the observation of agent $i$.
Details on the concrete observations, node features, and edge features used in this work are provided in Sec.~\ref{sec:rl-integration}.
Note that the invocations of $Q$ can be parallelized, i.e., $Q$ can be evaluated simultaneously for all neighbors.

\ifthenelse{\boolean{arxiv}}{
\begingroup
\definecolor{tabblue}{RGB}{31, 119, 180}
\definecolor{tabred}{RGB}{214, 39, 40}
\definecolor{tabgreen}{RGB}{44, 160, 44}
\definecolor{taborange}{RGB}{255, 127, 14}

\tikzset{
    node distance=1.5cm,
    node/.style={circle, draw, minimum size=.75cm},
    connection/.style={thick, gray},
    connection-wireless/.style={thick, gray, dotted},
    entangled/.style={line width=.5pt, tabblue, decorate, decoration={snake, amplitude=2pt, segment length=7pt}},
    entangled-strong/.style={line width=.5pt, tabblue, decorate, decoration={snake, amplitude=2pt, segment length=7pt}, double, double distance=1pt},
    agent/.style={rectangle, fill=taborange, minimum size=0.3cm},
    path/.style={line width=2.5pt, taborange, opacity=0.4},
    message/.style={->, thick, tabgreen, shorten >=2pt, shorten <=2pt, bend left=15},
    message-label/.style={font=\tiny, tabgreen}
}

\def\posAOnex{0.9}
\def\posAOney{-1.2}

\def\posATwox{1.2}
\def\posATwoy{0}

\def\posAThreex{2.4}
\def\posAThreey{0}

\def\posAFourx{3.9}
\def\posAFoury{-1.2}

\def\posASixx{2.4}
\def\posASixy{-1.2}

\def\posASevenx{3.6}
\def\posASeveny{0}

\def\posSOnex{2.4}
\def\posSOney{1.5}

\def\posPAOnex{0.4}
\def\posPAOney{-1.2}
\def\posPATwox{0.8}
\def\posPATwoy{0}
\def\posPASx{2.05}
\def\posPASy{1.5}
\def\posPAFourx{3.4}
\def\posPAFoury{-1.2}

\begin{figure*}[h]
    \centering
    \begin{minipage}{\textwidth}
        \centering
        \subfloat[Start of path planning\label{subfig:phases_first}]{
            \includegraphics{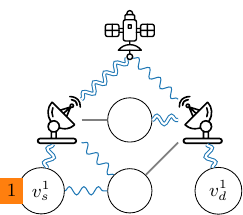}
        }
        \hspace{0.25cm}
        \subfloat[After first agent decision\label{subfig:phases_second}]{
            \includegraphics{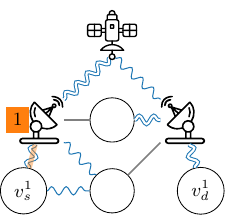}
        }
        \hspace{0.25cm}
        \subfloat[Agent moves to satellite\label{subfig:phases_third}]{
            \includegraphics{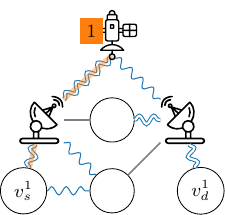}
        }
        \hspace{0.25cm}
        \subfloat[Finished path planning\label{subfig:phases_fourth}]{
            \includegraphics{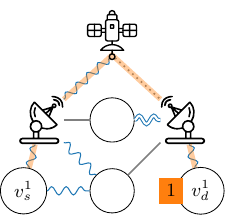}
        }
    \end{minipage}
    \caption{Visualization of path-planning in satellite-assisted quantum networks. The quantum repeaters are depicted as the nodes of the graph, which are connected via fiber. Dedicated ground stations have both fiber and satellite connections and serve as hubs between the topologies. If elementary links are available between repeaters, each blue wavy line represents one of these links. The agent is represented as an orange rectangle, which moves across the network during planning. The planned path is depicted as an orange line. Once the path is planned, the elementary links are consumed to create an end-to-end entanglement.}\label{fig:phases}
\end{figure*}
\endgroup
}{

}

\subsubsection{Improving the expressiveness of \glspl*{gnn}}
In related works, each repeater is typically modeled as a node in the GNN.
However, to make routing decisions, the focus is on edge-level predictions (see \eqref{eq:action-selection}).
This may require $Q$ to derive edge information from the node representations $h^v_{t+1}$ and $h^w_t$.
When assuming node embeddings of fixed size, they cannot encode features from an arbitrary number of incident edges without loss of information.
This can harm the predictive performance of the model.
Recent works found that edge-level predictions can be improved substantially by leveraging \emph{line graphs}~\cite{cai2021}.

A line graph transforms each edge $e\in E$ in the base graph to a node.
Two different line graph nodes $e_1 \neq e_2$ are connected by a line graph edge iff they represent adjacent edges in the base graph, i.e., $\exists v \in V:\: v\in e_1 \land v \in e_2$.
The \gls*{gnn} is then simply applied to the line graph.
To make edge-level predictions for an edge $e$ in the base graph, now only the node embedding $h^e$ of the corresponding node in the line graph needs to be considered as input to the scoring function $Q$.
Similarly, it has been shown that transformer models operating on node pairs instead of nodes can achieve higher expressiveness and better empirical performance~\cite{muller2024}.

The information flow between nodes in \gls*{gnn} models is typically bidirectional, which can lead to an over-homogenization of node embeddings.
Pahng and Hormoz~\cite{pahng2025} propose to explicitly learn the direction of the information flow per edge and show that this improves the expressiveness and empirical performance.

\subsubsection{Directed Line \gls*{gnn}}
\label{sec:directed-line-graph-gnn}
To combine the expected benefits of line graph \glspl*{gnn} and a directed information flow, we propose to apply the \gls*{gnn} on the \emph{directed line graph}, which is constructed as follows.
For each undirected edge $\{v, w\} \in E$ in the base graph $G$, the directed line graph has two nodes $a = (v, w),\, a^\prime = (w, v)$, representing the information flow in each direction.
We denote the set of line graph nodes that represent directed edges in the base graph $\vec E$.
Two line graph nodes $a, b \in \vec E$ are connected with a \emph{directed line graph edge} $(a, b) \in \vec A$ if there exists a common node $v\in V$ in the base graph such that $\exists u, w \in V:\: a = (u, v) \land b = (v, w)$, i.e., when $a$ is an incoming edge to $v$ and $b$ is an outgoing edge from $v$. 
We call the resulting directed graph $\vec G = (\vec E,\, \vec A)$ the directed line graph of $G$.

When applying the \gls*{gnn} architecture from \eqref{eq:message-passing} to the directed line graph, embedding information is now effectively managed for directed edges $(u, v) \in \vec E$, edge- and node-level input features are switched, and information is now forwarded backwards along the directed edges of the line graph.

Distributed execution of the resulting model is still possible.
However, as edges have no computational capacity in practice, the edge embeddings must be managed by nodes in the network.
A naive strategy is to let node $u$ manage the edge embeddings of all outgoing edges $(u, w) \in \vec E$.
Note that, when applying \eqref{eq:message-passing} to the base graph, each node manages a node embedding of fixed dimensionality.
With the directed line graph model, each node $u$ now manages embedding information of varying dimensionality, depending on the number of incident edges.

\subsection{Integration into Reinforcement Learning}
\label{sec:rl-integration}
An end-to-end entanglement request consists of a unique identifier $i$, a source-destination pair $(v^i_s, v^i_d)$, and a \gls*{ttl}.
This request serves as the basis for the path-planning of our reinforcement learning agent.
Similar to related work~\cite{meuser2025}, SatQNet plans paths iteratively.
However, in contrast to related work, the path-planning may consider links to satellites to cover large distances, as visualized in Fig.~\ref{fig:phases}.
As depicted in Fig.~\ref{subfig:phases_first}, agent $i$ is instantiated at its source quantum repeater $v^i_s$ and aims to move towards its destination repeater $v^i_d$.
At each environment step, the agent is located at exactly one repeater, as depicted in Figs.~\ref{subfig:phases_second} and \ref{subfig:phases_third}.
The agent can choose any of the physical connections available as its next step along the path.
The operation of the agent is decoupled from the entanglement swap operation, such that unnecessary decays of created entanglements are avoided.
Only once a path has been planned, as shown in Fig.~\ref{subfig:phases_fourth}, an entanglement swap is performed on all repeaters along the path to create an end-to-end entanglement.

\subsubsection{Observation Space}
The observation space contains information about the end-to-end entanglement request associated with the agent and about the neighboring quantum repeaters and the connection to them.

For a request, the agent observes the length of its current path to track the number of steps that remain until the \gls*{ttl} expires and path-planning fails.

Recall that at each step, the agent is located at a quantum repeater.
For each neighboring quantum repeater $v_j$, the agent observes its swap probability and its role in the request.
The agent observes a binary feature that indicates whether the neighbor $v_j$ is the destination $v^i_d$ of agent $i$, i.e., $v^i_d = v_j$.
In addition, it observes whether $v_j$ is in the same ground cluster as $v^i_d$.
This topological information supports the path-planning of the agent in the satellite topology with multiple clusters.
For each outgoing connection, the number of available elementary links on the connecting optical link and the highest fidelity of the available elementary links are observed.

Note that the agent neither directly observes the global location of the destination nor does it receive any shortest-path heuristics.
Instead, these properties are learned by the quantum repeaters.
In addition to handcrafted observations, the agent receives the learned \gls*{gnn} embedding information available to the quantum repeater it is located on (compare \eqref{eq:action-selection}).
When using the directed line \gls*{gnn} proposed in Sec.~\ref{sec:directed-line-graph-gnn}, the agent observes the edge embeddings of all outgoing links.

\subsubsection{Node and Edge Features}
For each simultaneous source-destination pair $i$, a separate instance of the GNN is required.
While this introduces overhead for each additional source-destination pair, previous work has shown that the required resources are relatively low~\cite{meuser2025}.
As input to each instance of the \gls*{gnn} for source-destination pair $i$, node and edge features are required.
The node features describe properties of the quantum repeater, specifically its swap probability and its role in a specific request $i$.
While the swap probability is independent of $i$, the role depends on $i$ and is divided into two aspects, similar to the agent's observation.
The first is a binary feature that indicates whether node $v$ is the destination $v^i_d$ of agent $i$, i.e., $v^i_d = v$.
In addition, we encode whether $v$ is in the same ground cluster as $v^i_d$.
The edge features describe properties of a connection between two quantum repeaters, specifically the number and quality of available elementary links as well as the probability of a successful establishment of a new elementary link.

\subsubsection{Action Space}
The action space of the agent $i$ depends on the quantum repeater $v^i_c$ that it is currently associated with, as shown in Fig.~\ref{fig:phases}.
The size of the action space matches the degree $\left|\mathcal{N}(v^i_c)\right|$ of this quantum repeater $v^i_c$, and the action $a^i_t \in \mathcal{N}(v^i_c)$ corresponds to the next hop.
The agent may choose any outgoing connection of the repeater, which will move it to the neighbor associated with this connection as shown in Fig.~\ref{subfig:phases_second}.
This adds a reservation to the connection, which is not bound to any specific elementary link but always reserves the elementary link with the highest fidelity.
The reservations are removed once the path-planning is completed and the elementary links are used to create an end-to-end entanglement.
Without such reservations, resource contention is a significant issue in multi-agent scenarios.

\subsubsection{Objective and Evaluation Function}
The objective of agent $i$ is to select a path to its destination $d^i$ that maximizes the expected end-to-end fidelity while limiting resource consumption, thereby supporting a larger total number of established entanglements.
As the end-to-end fidelity is only known upon completion of the path, related work~\cite{meuser2025} utilizes a sparse reward function, where rewards are given upon completion of a path based on its end-to-end fidelity.

The end-to-end fidelity $F(\tau^i_{0:T})$ depends on the trajectory of nodes $\tau^i_{0:T} = (v_0,\, v_1,\, \dots,\, v_T)$ visited by agent $i$ from step $t=0$ to step $T$.
It is determined by recursive swap operations between all intermediate neighbor nodes as
\begin{equation}
F(\tau^i_{t:T}) = 
    \begin{cases}
        F(\tau^i_{t+1:T}) & \text{if } v_t = v_{t+1},\\[6pt]
        f_{\{v_t, v_T\}} & \text{if } \{v_t, v_T\} \in E,\\[6pt]
        \operatorname{swap}(f_{\{v_t, v_{t+1}\}}, F(\tau^i_{t+1:T})) & \text{otherwise,}
    \end{cases}
\label{eq:end-to-end-fidelity}
\end{equation}

where the function $\operatorname{swap}$ models the entanglement swap operation (see Section~\ref{sec:entanglement-swap}) and $f_{\{u, w\}}$ is the fidelity of a direct entanglement from node $u$ to its neighbor $w$.

The goal of reinforcement learning methods is to find an optimal policy that maximizes the expected discounted return $G^i_t = \sum_{k={t+1}}^T\gamma^{k-t-1} r^i_k$.
Given a sparse reward $r^i_T = F(\tau^i_{0:T})$ only at the end of a path if an end-to-end entanglement to the destination could be established, the discounted return for any intermediate step $t\in\{0, 1, \dots, T-1\}$ reduces to $G^i_t = \gamma^{T-t-1} F(\tau^i_{0:T})$.
In our quantum routing environment, the discount factor $\gamma \in (0,1]$ controls the trade-off between entanglement quality and resource consumption. 
A larger $\gamma$ results in a prioritization of high-fidelity elementary links, yielding higher-quality end-to-end entanglement at the expense of longer paths and thus increased resource usage, which may reduce the overall entanglement generation rate.

\begin{figure*}[t]
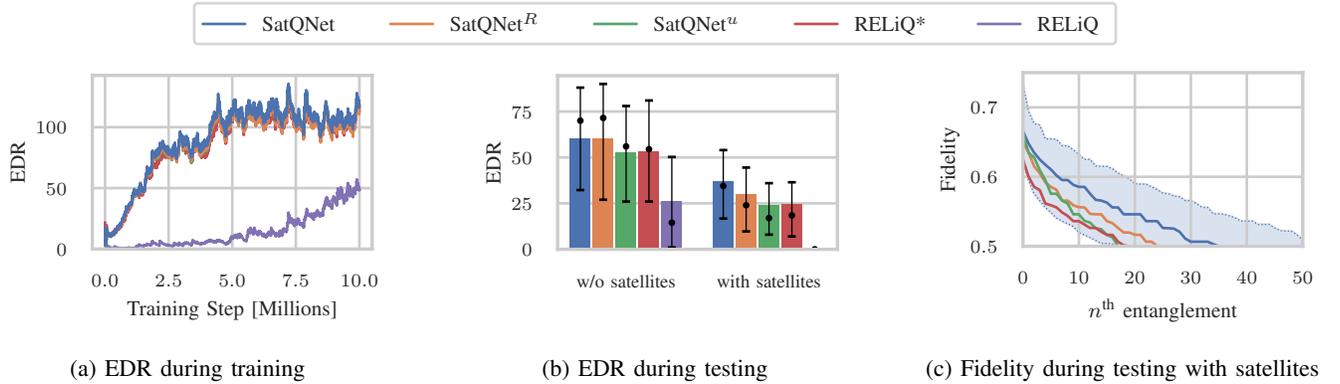

    \centering
    \input{gfx/eval/no_legend/train/average_episode_packets/legend_training.pgf}\\
    \begin{minipage}{\textwidth}
    \centering
    \subfloat[EDR during training\label{subfig:training_edr}]{
        \input{gfx/eval/no_legend/train/average_episode_packets/line___training.pgf}
    }\hfill
    \subfloat[EDR during testing\label{subfig:testing_edr}]{
        \input{gfx/eval/no_legend/average_episode_packets/bar___network_islands_trained_only.pgf}
    }\hfill
    \subfloat[Fidelity during testing with satellites\label{subfig:testing_fidelity}]{
        \input{gfx/eval/no_legend/fidelities/behavior___fidelity_trained_only.pgf}
    }
    \end{minipage}
    \caption{Performance of SatQNet compared to learning-based approaches.}
    \label{fig:training_visualization}
\end{figure*}

In theory, this framework allows learning optimal paths in any network. 
In practice, however, we found that it is challenging to estimate the quality of long paths, resulting in poor decisions and hence the selection of suboptimal paths.
Even if a path from the source to the destination node is found, when elementary links along the path have low fidelity, the end-to-end fidelity may become zero, and agents thus may not receive any meaningful feedback to improve their policy.

To address this issue, one would like to define a dense reward function that assigns rewards to each intermediate action.
Unfortunately, \eqref{eq:end-to-end-fidelity} cannot easily be decomposed into intermediate rewards, as the $\operatorname{swap}$ operation does not behave additively (as the rewards in the discounted return) and because the end-to-end fidelity ultimately depends on all elementary links in the chosen path to the destination.

In this work, as an alternative to a dense reward function, we propose to assess the quality of the trajectory $\tau^i_{t:T}$ of agent $i$ from step $t$ to step $T$ with the following \emph{evaluation function}
\begin{equation}\label{eq:return}
    E^i(\tau^i_{t:T}) =
    \begin{cases}
        \gamma^{T-1-t} F(\tau^i_{t:T}) & \text{if } v_T = d^i,\\[6pt]
        0 & \text{otherwise}.
    \end{cases}
\end{equation}
Intuitively, it evaluates the quality of the subpath from the current node $v_t$ to the final node $v_T$, assuming that the current node was the start node.
Given a complete trajectory $\tau^i_{0:T}$, this yields qualities for each intermediate step $t < T$.
Note that for $t=0$, the evaluation function is equivalent to the discounted return.
However, for $t > 0$, and particularly when $v_t$ is close to the destination $d^i$, the quality returned by the evaluation function may be positive, even if the entanglement from the source node $v_0$ to the destination could not be established.

The evaluation function provides a more fine-grained learning target than the sparse reward function, but deviates from common reinforcement learning frameworks.
Reward functions are usually defined on a single state transition, and recent related works investigate non-Markovian rewards that depend on the history of past states and actions~\cite{gaon2020, lin2025}.
In our case, however, the evaluation function depends on \emph{future states and actions} and can thus only be computed upon episode completion. 
While our work focuses on the practical effect of this design decision on quantum routing, this may also provide an interesting opportunity for future basic research.

Agents are trained by iteratively approximating the qualities returned by the evaluation function and acting $\epsilon$-greedily upon the predicted qualities.
We leverage an experience replay memory, as in deep $Q$-learning~\cite{mnih2015}, to improve stability.
However, instead of the temporal-difference method, we simply minimize the mean squared error between the predicted qualities and those obtained via the evaluation function.

\section{Evaluation}
\label{evaluation}

We analyze the performance of SatQNet in diverse scenarios and settings using the evaluation framework developed in \cite{meuser2025}, which we extend with satellites and ground stations.
The satellite topology is based on the Starlink topology, which we obtained using CelesTrak\footnote{\url{https://celestrak.org/}}.
In that version, there are $9307$ satellites available, which is slightly lower than the current estimate of approximately $10000$ satellites.
During the simulation, the satellites move using predictions from the Simplified General Perturbations (SGP4) model.
The ground topology and the satellite topology are connected using ground stations, which have both a fiber connection and a wireless link to exchange data and create elementary links with satellites.

We evaluate SatQNet against seven representative entanglement-routing baselines: Q-PATH and Q-LEAP~\cite{li2022}, MGER, LBER, and NoNLBER~\cite{chakraborty2019}, AER~\cite{xiong2023}, and RELiQ~\cite{meuser2025}.
These algorithms were selected to cover different routing strategies and information assumptions. MGER, LBER, NoNLBER, and RELiQ are decentralized and rely only on local quantum topology information.
In contrast, Q-LEAP, Q-PATH, and AER rely on centralized coordination with global topology information, which in our simulation is collected through a dedicated monitoring framework.
Especially for large topologies, centralized methods can suffer from stale information about distant quantum links.
RELiQ is also decentralized and learning-based, and is therefore the most closely related baseline to our approach.

To compare the performance of the approaches, we utilized the Entanglement Distribution Rate (EDR) as main metric, which corresponds to the number of end-to-end entanglement created over the course of one episode.
In addition, the fidelity of the created entanglements is used to assess their usability for quantum applications.
If the fidelity is not mentioned explicitly, the fidelity of entanglements created by SatQNet is similar or higher to those of the other approaches.

To assess whether observed performance differences are significant, we use one-sided paired permutation tests based on the mean values ($H_0$: baseline $\geq$ SatQNet; $H_1$: SatQNet $>$ baseline), with \num{100000} random permutations per test.
Such tests are effective in analyzing classifier performance~\cite{ojala2010,hastie2009}.
When we compare SatQNet against multiple baselines, the p-values are adjusted with the Holm--Bonferroni method.
Pairwise comparisons (e.g., SatQNet vs.\ $\text{SatQNet}^u$) are reported without that adjustment.

\begin{figure*}[t]
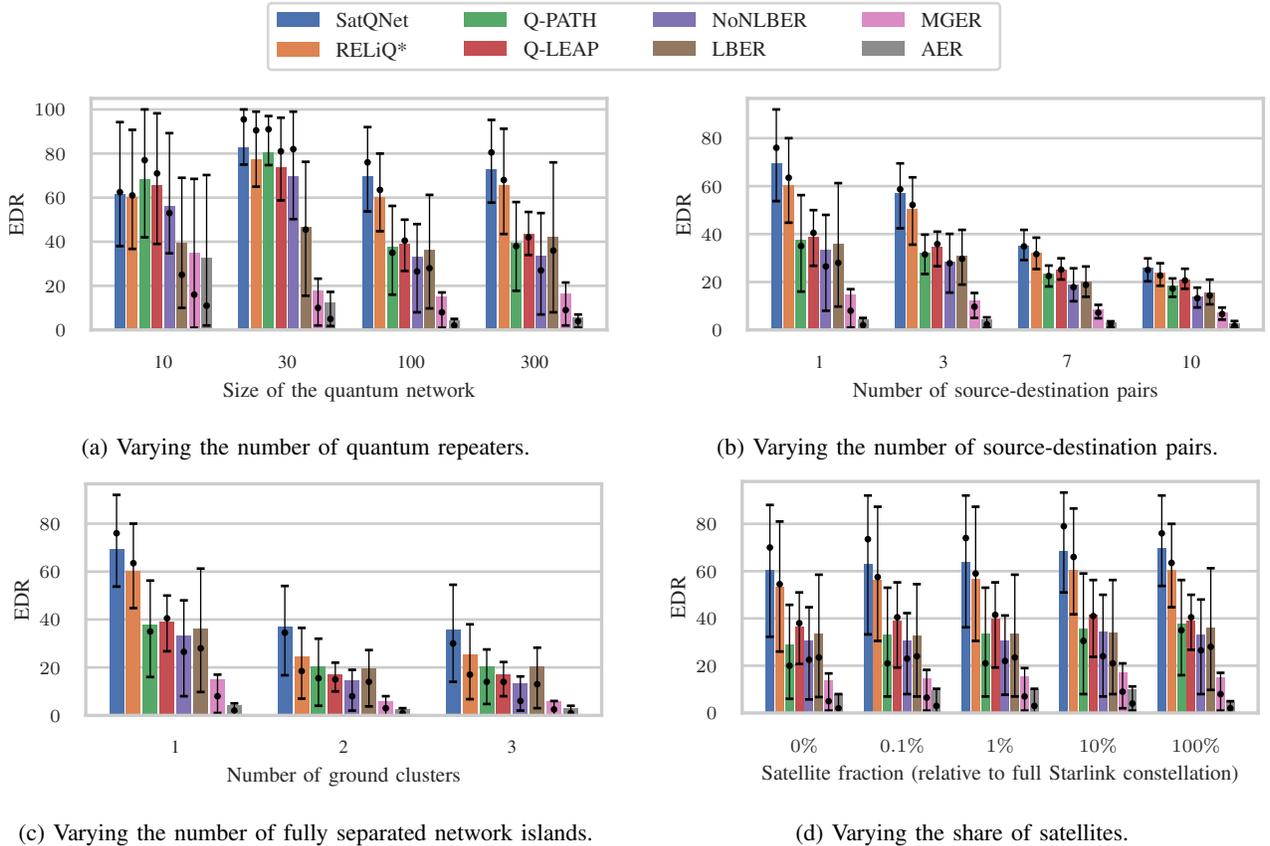

    \centering
    \input{gfx/eval/no_legend/average_episode_packets/legend.pgf}\\
    \subfloat[Varying the number of quantum repeaters.\label{fig:packets_per_network_size}]{\input{gfx/eval/no_legend/average_episode_packets/bar___n_router.pgf}}\quad
    \subfloat[Varying the number of source-destination pairs.\label{fig:packets_per_source_destination_pairs}]{\input{gfx/eval/no_legend/average_episode_packets/bar___n_data.pgf}}\\
    \subfloat[Varying the number of fully separated network islands.\label{fig:packets_per_network_islands}]{\input{gfx/eval/no_legend/average_episode_packets/bar___network_islands.pgf}}\quad
    \subfloat[Varying the share of satellites.\label{fig:packets_per_share_of_satellites}]{\input{gfx/eval/no_legend/average_episode_packets/bar___satellite_share.pgf}}
    \caption{Entanglement Distribution Rate (EDR) based on the scale and topology of the quantum network.}\label{fig:network_topology}
\end{figure*}

\subsection{Learning-based Approaches and Ablations}
We compared SatQNet against other learning-based approaches, which use local observations and message exchange to make routing decisions.
We utilized $\epsilon_\text{decay} = 0.9999$, $\gamma = 0.95$, a learning rate of $0.0005$, a sequence length of $20$, a mini-batch size of $32$, and a replay buffer with \num{100000} entries.
Each approach was trained in episodes of \num{1000} steps each, up to a total of \num{10000000} steps.
During an episode, an end-to-end entanglement request is created every \SI{100}{\milli\second}.
We trained on a topology of two ground clusters, which are connected by satellites.

Learning approaches that are not able to generalize to new topologies are excluded here, as they cannot handle the dynamically changing topologies in our evaluation.
SatQNet is our proposed solution based on directed line graphs and the evaluation function described in \eqref{eq:return}, while $\text{SatQNet}^R$ utilizes the reward proposed in RELiQ~\cite{meuser2025} and $\text{SatQNet}^u$ utilizes only an undirected line graph.
As related learning-based baselines, we compare against RELiQ~\cite{meuser2025} and a slightly improved version of it, which we label RELiQ*.
RELiQ* is an extension of RELiQ developed to make it applicable to satellite-assisted networks and relies on node observations that are independent of the connectivity of quantum repeaters by using a similar mechanism as described in Sec.~\ref{subsubsec:dynamic_node_degrees}.
It can be seen in Fig.~\ref{subfig:training_edr} that RELiQ struggles to converge in satellite-assisted networks due to the high and strongly fluctuating node degree.
RELiQ* solves this issue and remains competitive even in satellite-assisted networks.
The three versions of SatQNet perform similarly during training, with only minor differences in the number of created end-to-end entanglements.
However, when evaluating the resulting models in Fig.~\ref{subfig:testing_edr}, we observe differences in their learned behavior.
We compare two settings, one in which no satellites are present, and another in which satellites are essential to connect two ground clusters each with $100$ ground repeaters.
The high node degree of satellite networks leads to performance reductions of RELiQ* in satellite-assisted topologies.
Similarly, when utilizing SatQNet based on a undirected line GNN, the performance reduces significantly in the satellite topology ($p < 0.001$).
In Fig.~\ref{subfig:testing_fidelity}, the fidelity of each approach is shown for the satellite-based topology.
While all approaches create a few high-fidelity entanglements, the fidelity of successive entanglements is consistently higher for SatQNet.

\begin{figure*}[t]
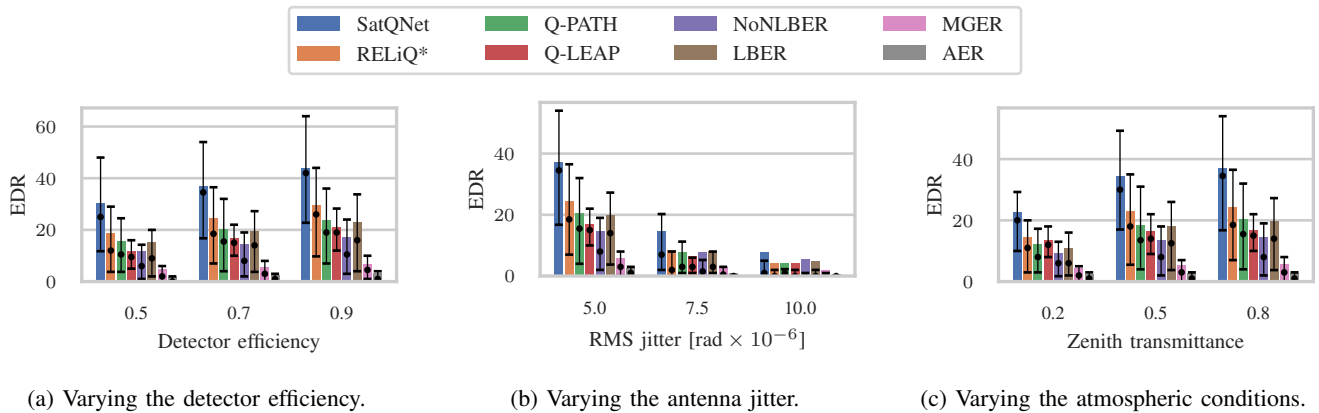

    \centering
    \input{gfx/eval/no_legend/average_episode_packets/legend.pgf}\\
    \begin{minipage}{\textwidth}
    \centering
    \subfloat[Varying the detector efficiency.\label{subfig:sat_detector_efficiency}]{
        \input{gfx/eval/no_legend/average_episode_packets/bar___sat_detector_efficiency.pgf}
    }\hfill
    \subfloat[Varying the antenna jitter.\label{subfig:sat_rms_jitter}]{
        \input{gfx/eval/no_legend/average_episode_packets/bar___sat_rms_jitter.pgf}
    }\hfill
    \subfloat[Varying the atmospheric conditions.\label{subfig:zenith_transmittance}]{
        \input{gfx/eval/no_legend/average_episode_packets/bar___zenith_transmittance.pgf}
    }
    \end{minipage}
    \caption{Influence of satellite parameters on the performance of the approaches.}
    \label{fig:sat_parameters}
\end{figure*}

\subsection{Results}
Fig.~\ref{fig:packets_per_network_size} displays the performance of the different routing algorithms for different sizes of the ground cluster.
In each plot, the bar height is the mean EDR over \num{100} episodes per approach.
Error bars span the 25th to 75th percentile across those episodes; a marker indicates the median.
Permutation tests use the per-episode EDR values as samples.
The performance of global-knowledge-based approaches Q-PATH and Q-LEAP is very high for small network sizes, as information about the underlying topology can still be obtained with almost no latency.
When increasing the number of ground nodes from $10$ to $30$, the performance of most approaches increases, as more resources are available for entanglement routing.
As network size increases further, the performance of Q-PATH and Q-LEAP diminishes drastically, such that even comparatively simple local-knowledge-based approaches become competitive.
Similar to SatQNet, RELiQ* performs well across all network sizes.
However, SatQNet achieves a higher EDR than RELiQ* for all evaluated network sizes, as its edge-based view of the network adapts better to the varying connectivity of quantum repeaters.
For a network with at least $30$ ground nodes, SatQNet outperforms every baseline; all seven comparisons are significant (Holm-corrected $p < 0.05$ for $30$ nodes, $p < 0.001$ for $100$ and $300$ nodes).

Fig.~\ref{fig:packets_per_source_destination_pairs} shows the performance per source-destination pair of all approaches when multiple source-destination pairs need to be served simultaneously.
While the EDR decreases for all approaches due to resource contention, it can be observed that SatQNet performs well in this multi-agent setting, even though only a single agent was considered during training.
Its performance is consistently higher than all baseline approaches (Holm-corrected $p < 0.001$), but the relative improvement reduces once the network becomes overloaded with requests.
This is especially impressive, as the global-knowledge-based approaches have a clear advantage due to the additional coordination that is possible during planning at a central instance.
From $7$ to $10$ simultaneous source-destination pairs, the total number of established entanglements stagnates, as the generation of elementary links becomes the limiting factor.
However, we observed that the fidelity of the created entanglements is higher for Q-PATH and Q-LEAP compared to SatQNet when resource contention is high.

Fig.~\ref{fig:packets_per_network_islands} shows the effect of completely separated ground clusters.
If more than one ground cluster is simulated, the start and destination are always placed in different ground clusters to enforce the usage of satellites.
It is evident that the performance of all approaches decreases when satellite-based links must be used, as the relatively high distance between satellites causes low elementary link generation rates.
However, the results for $2$ and $3$ ground clusters are very similar for all approaches, highlighting that they can correctly identify the destination cluster.
In all cases, SatQNet significantly outperforms all other approaches (Holm-corrected $p < 0.001$).

Fig.~\ref{fig:packets_per_share_of_satellites} shows the influence of the share of satellites for a single ground cluster.
Thus, entanglements can always be created using ground-only paths.
Even when the satellite share is small, the performance of all approaches increases slightly.
This shows that ground-based quantum networks already benefit from a comparatively small number of satellites in orbit.
In all configurations, SatQNet outperforms all other approaches significantly (Holm-corrected $p < 0.001$).
This highlights the adaptability of SatQNet to different conditions.

As already discussed in Sec.~\ref{sec:link-success-probabilities}, the parameters of the satellite model have a strong influence on the success probability for establishing elementary links, thus limiting the number of end-to-end entanglements that can be created.
Fig.~\ref{fig:sat_parameters} shows the influence of three important parameters, the efficiency of the detector, the jitter of the antenna, and the zenith transmittance.
The results are generated in a setting with two ground clusters to enforce the utilization of satellite links.
As shown in Fig.~\ref{subfig:sat_detector_efficiency}, the quality of the detector has a strong influence on the results.
All approaches adapt well to the increasing efficiency and create more end-to-end entanglements.
The jitter of the antenna has the opposite effect. As shown in Fig.~\ref{subfig:sat_rms_jitter}, high antenna jitter leads to a drastic drop in the performance of all approaches, highlighting the need for precise antenna control in satellite-assisted quantum networks.
Zenith transmittance, as a measure of photon absorption in the atmosphere, also has a significant influence, with all approaches adapting to changing conditions.
In all satellite scenarios, SatQNet adapts well to changing conditions and maintains a high EDR, outperforming all other approaches by a significant margin (Holm-corrected $p < 0.001$).

\begin{figure}[thb]
    \centering
    \input{gfx/eval/legend/average_episode_packets/colorbar.pgf}\\
    
    \subfloat[Q-PATH\label{subfig:map_qpath}]{
        \includegraphics[width=.47\linewidth]{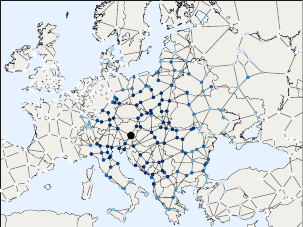}
    }\hfill
    \subfloat[SatQNet\label{subfig:map_satqnet}]{
        \includegraphics[width=.47\linewidth]{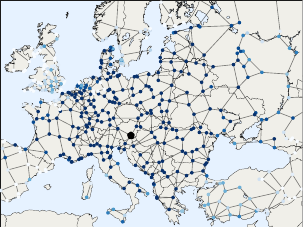}
    }
    \caption{Connectivity of Klagenfurt to other European cities.}
    \label{fig:map}
\end{figure}

\subsection{Application to real-world backbone networks}
In this section, we analyze the performance of SatQNet on real-world topologies.
Even though SatQNet is trained on random topologies, applying it to real-world backbones without retraining provides a practical test of its adaptability to unseen graph structures.
Fig.~\ref{fig:map} shows the achieved end-to-end entanglement rate from Klagenfurt (Austria) to major European cities when satellite assistance is enabled.
For this evaluation, up to $10$ entanglements are created between Klagenfurt and any other city in Europe, and the results are reported as a success rate.
Note that performing episodes of $1000$ steps as for the other experiments was not desirable due to the high number of European cities ($>500$), so we used only $100$ steps to reduce the computational load.
Klagenfurt is marked with a black pentagon.
The node colors indicate the percentage of successfully created end-to-end entanglements, ranging from $0\%$ (white) to $100\%$ (dark blue).
We observe that SatQNet achieves a higher success rate than Q-PATH for $97\%$ of destination cities.
For the permutation test, we use the success rate to all destination cities as samples.
When comparing the overall success rates between SatQNet and Q-PATH, with each city contributing one sample, the Europe-wide difference is statistically significant ($p < 0.001$).
While both approaches perform similarly in parts of central Europe, their performance diverges with increasing distance, where SatQNet shows clear gains, including in northern Germany, Denmark, France, and Turkey.
These results highlight that SatQNet generalizes well beyond its training distribution and remains effective on realistic large-scale backbone topologies.

\section{Conclusion}
\label{conclusion}

We have presented SatQNet, a decentralized reinforcement learning-based approach relying on a directed line \gls*{gnn} for entanglement routing in satellite-assisted quantum networks.
Existing approaches often rely on global information about the quantum topology or assume a static physical topology, which limits their scalability and their ability to route efficiently in large-scale, highly dynamic quantum networks.
SatQNet addresses this challenge by learning graph representations from local observations.
In contrast to existing learning-based approaches that rely on node embeddings, SatQNet maintains edge embeddings at each node, allowing it to adapt dynamically to the varying connectivity of each repeater.

The evaluation on random graphs shows that SatQNet significantly outperforms learning-based approaches relying on node observations in all evaluated settings.
Additionally, SatQNet outperforms non-learning-based approaches in large networks, where the latency of collected global information becomes prohibitive.
It also handles resource scarcity when multiple source-destination pairs compete for resources, and achieves clear gains when the end-to-end entanglement requires satellite-assisted paths.
The advantage also persists under changes in satellite parameters, indicating that the learned policy adapts well to varying physical conditions.

Beyond synthetic topologies, SatQNet generalizes to a real-world European backbone without retraining, where it achieves a higher success rate than Q-PATH for $97\%$ of destination cities.
Together, these results support the central claim of this work: decentralized, edge-centric learning is a scalable and effective approach to entanglement routing in highly dynamic satellite-assisted quantum networks.

In future work, we plan to improve the resource efficiency of SatQNet by adaptively choosing the message size depending on the expected impact on the reinforcement learning agent and investigate strategies for asynchronous environments.

\section*{Acknowledgment}
During the preparation of this manuscript, the authors used \textit{Generative AI} to improve the spelling, grammar, and readability of the text. After using this tool, the authors reviewed and edited the content as needed and take full responsibility for the final version of the manuscript.

\balance

\bibliographystyle{IEEEtran}
\bibliography{bibliography}

\end{document}